\documentclass[apj]{emulateapj}
\newcommand{\kms}{km~s$^{-1}$}
\newcommand{\vsini}{$v\sin i$}

\newcommand{\msun}{$M_{\sun}$}

\newcommand{\ali}{$A$(Li)}
\newcommand{\cratio}{$^{12}$C/$^{13}$C}
\newcommand{\thestar}{2M19411367+4003382}
\newcommand{\thecluster}{NGC~6819}
\newcommand{\teff}{$T_{\rm eff}$}
\shorttitle{}
\shortauthors{Carlberg et al.}

\begin{document}

\title{The Puzzling Li-rich Red Giant Associated with NGC 6819}
\shorttitle{Li-rich Red Giant Associated with NGC~6819}

\author{Joleen K. Carlberg\altaffilmark{1, 23}, Verne V. Smith\altaffilmark{2},   Katia Cunha\altaffilmark{3,22}, Steven R. Majewski\altaffilmark{4},  Szabolcs~M{\'e}sz{\'a}ros\altaffilmark{5,6}, Matthew Shetrone\altaffilmark{7}, Carlos Allende Prieto\altaffilmark{8,9}, Dmitry Bizyaev\altaffilmark{10,11}, Keivan G. Stassun\altaffilmark{12, 13}, 
Scott W. Fleming\altaffilmark{14, 15}, Gail Zasowski\altaffilmark{16, 17}, Fred Hearty\altaffilmark{18}, David L. Nidever\altaffilmark{19}, Donald P. Schneider\altaffilmark{18,20}, Jon A. Holtzman\altaffilmark{11}, Peter M. Frinchaboy\altaffilmark{21}} 

\altaffiltext{1}{Department of Terrestrial Magnetism, Carnegie Institution of Washington, 
5241 Broad Branch Road, NW, Washington DC 20015, USA}
\altaffiltext{2}{National Optical Astronomy Observatory, 950 North Cherry Avenue, Tucson, AZ 85719, USA}
\altaffiltext{3}{Observat\'orio Nacional, Rua General Jos\'e Cristino, 77, 20921-400 S\~ao Crist\'ov\~ao, Rio de Janeiro, RJ, Brazil}
\altaffiltext{4}{Department of Astronomy, University of Virginia, Charlottesville, VA 22904, USA}
\altaffiltext{5}{ELTE Gothard Astrophysical Observatory, H-9704 Szombathely, Szent Imre herceg st. 112, Hungary}
\altaffiltext{6}{Department of Astronomy, Indiana University, Bloomington, IN 47405, USA;}
\altaffiltext{7}{McDonald Observatory, University of Texas at Austin, McDonald Observatory, TX 79734, USA}
\altaffiltext{8}{Instituto de Astrof\'isica de Canarias, C/V\'ia L\'actea, s/n, E-38200, La Laguna, Tenerife, Spain}
\altaffiltext{9}{Departamento de Astrof\'isica, Universidad de La Laguna, E-38206, La Laguna, Tenerife, Spain}
\altaffiltext{10}{Apache Point Observatory, P.O. Box 59, Sunspot, NM 88349, USA}
\altaffiltext{11}{Department of Astronomy, MSC 4500, New Mexico State University, P.O. Box 30001, Las Cruces, NM 88003, USA}
\altaffiltext{12}{Department of Physics and Astronomy, Vanderbilt University, VU Station 1807, Nashville, TN 37235, USA}
\altaffiltext{13}{Department of Physics, Fisk University, 1000 17th Avenue North, Nashville, TN 37208, USA}
\altaffiltext{14}{Space Telescope Science Institute, 3700 San Martin Drive, Baltimore, MD, 21218 USA}
\altaffiltext{15}{Computer Sciences Corporation, 3700 San Martin Drive, Baltimore, MD, 21218 USA}
\altaffiltext{16}{Center for Astrophysical Sciences, Department of Physics and Astronomy, Johns Hopkins University, 3400 North Charles Street, Baltimore, MD 21218, USA}
\altaffiltext{17}{NSF Astronomy and Astrophysics Postdoctoral Fellow}
\altaffiltext{18}{Department of Astronomy and Astrophysics, The Pennsylvania State University, 525 Davey Lab, University Park PA 16802, USA}
\altaffiltext{19}{Department of Astronomy, University of Michigan, Ann Arbor, MI, 48104, USA}
\altaffiltext{20}{Institute for Gravitation and the Cosmos, The Pennsylvania State University, University Park, PA 16802, USA}
\altaffiltext{21}{Department of Physics \& Astronomy, Texas Christian University, Fort Worth, TX 76129, USA}
\altaffiltext{22}{Steward Observatory, University of Arizona, Tucson, AZ 85719, USA}
\altaffiltext{23}{NASA Postdoctoral Program Fellow, joleen.k.carlberg@nasa.gov}

\begin{abstract}
A Li-rich red giant star (2M19411367+4003382)   recently discovered in the direction of NGC~6819  belongs to the rare subset  of Li-rich stars  that have   not yet evolved to the luminosity bump, an evolutionary stage where models predict Li can be replenished. 
The currently favored model to explain Li enhancement in first-ascent red giants like 2M19411367+4003382 requires deep mixing into the stellar interior. 
Testing this model requires a measurement of \cratio, which is possible to obtain from APOGEE spectra. 
However, the Li-rich star  also has  abnormal asteroseismic properties that call into question its membership in the cluster, even though its  radial velocity and location on color-magnitude diagrams are consistent with membership.
To address these puzzles, we have measured a wide array of abundances in the Li-rich star and three comparison stars using spectra taken as part of the APOGEE survey to determine the degree of stellar mixing, address the question of membership, and measure the surface gravity.  We confirm that the Li-rich star is a red giant with the same overall chemistry as the other cluster giants.
However, its $\log g$ is significantly lower, consistent  with the asteroseismology results and suggestive of a very low mass if the star is indeed a cluster member. 
Regardless of the cluster membership, the \cratio\ and C/N  ratios of the Li-rich star are  consistent with standard first dredge-up, indicating  that Li dilution has already occurred, and inconsistent with  internal Li enrichment scenarios that require deep mixing.
\end{abstract}

\section{Introduction}
\label{sec:intro}
\citet[][hereafter AT13]{2013ApJ...767L..19A}   recently reported the discovery of a Li-rich red giant (RG) star in NGC 6819. This star, 2M19411367+4003382, 
has    \ali$\sim 2.3$~dex ([Li/H]$\sim$1.3) and is unusual even within the class of rare Li-rich giants.  
The star's current position on a color-magnitude diagram (CMD) suggests that it has evolved beyond the Li dilution phase of  first dredge-up (FDU), and, indeed, 
the other NGC~6819 RGs at similar magnitudes have   \ali$<0.7$~dex (AT13). What makes this Li-rich star so unusual is that its CMD position is clearly below and blueward of the luminosity bump, the evolutionary stage where models  have demonstrated that newly synthesized  Li can be circulated to the stellar envelope, resulting in a brief stage of Li-richness (see, e.g., \citealt{Charbonnel:2000ud}, \citealt{eggleton08},  \citealt{denissenkov12}, hereafter D12).   Other  recent studies (e.g., \citealt{Kumar:2011jr} and \citealt{carlberg12}) have  identified Li-rich stars in the field that are too hot to be bump stars. The former study hypothesized that these stars are red clump (RC) stars that replenished Li during the He flash, and the low \cratio\ of many of these stars are consistent with  internal Li regeneration. The latter study favored  planetary engulfment because their stars  tend to be more rapidly rotating and have \cratio\ near normal FDU values.

The currently favored model for \thestar\ is the mixing model of D12, which can explain Li-rich giants below the luminosity bump. In addition to enriching the surface with Li, the mixing alters the star's evolution, making it follow a more extended ``retrograde'' evolutionary path down the red giant branch (RGB) compared to normal luminosity bump evolution.
In the discovery paper,  AT13  argued that \thestar's  Li abundance and position on the RGB  was most consistent with this model, but they also noted that a crucial test of this hypothesis is the measurement of \cratio\ to confirm whether the extra mixing has indeed occurred.

\begin{figure}[tb]
\includegraphics[width=0.5\textwidth]{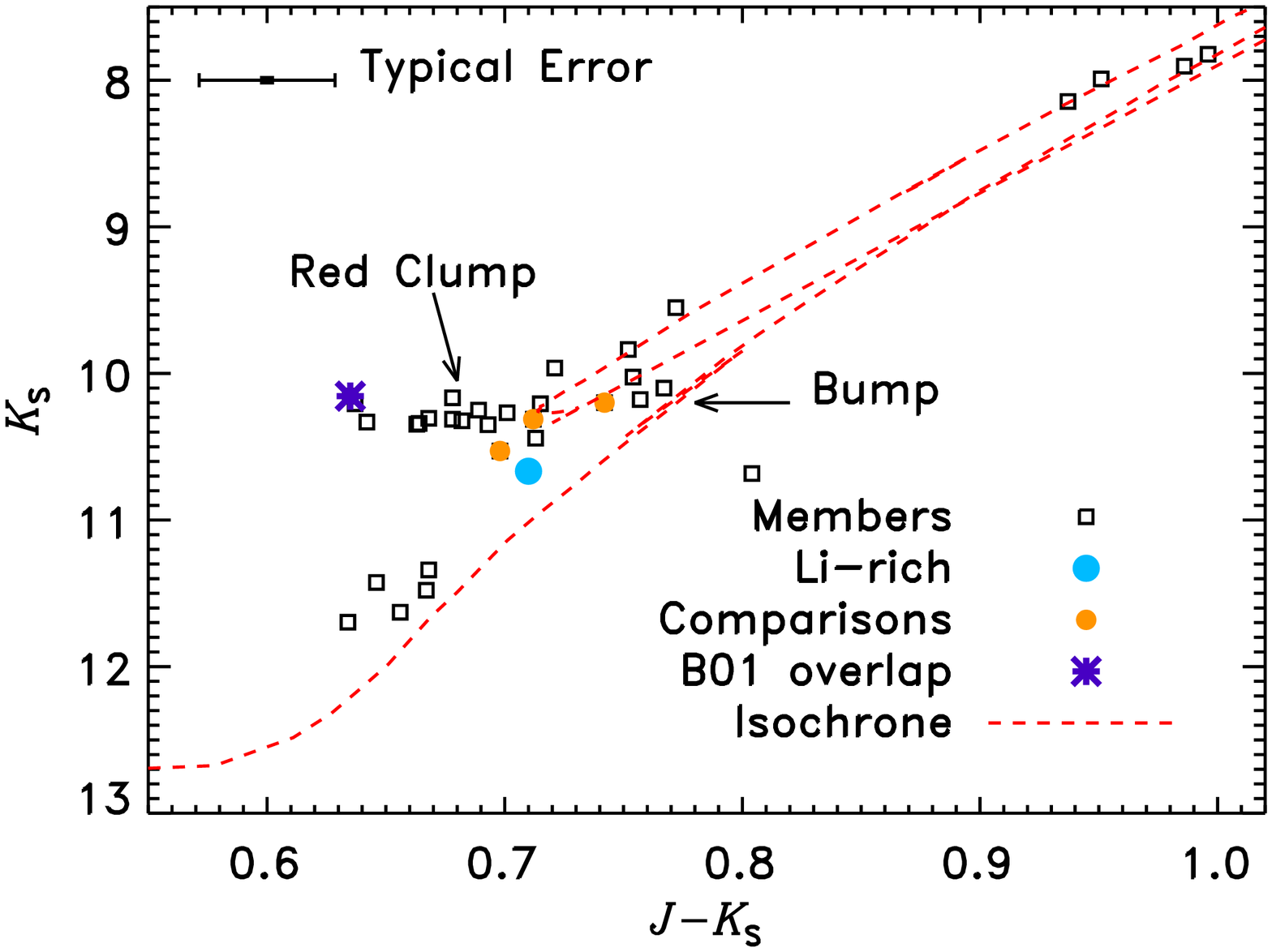} 
\includegraphics[width=0.5\textwidth]{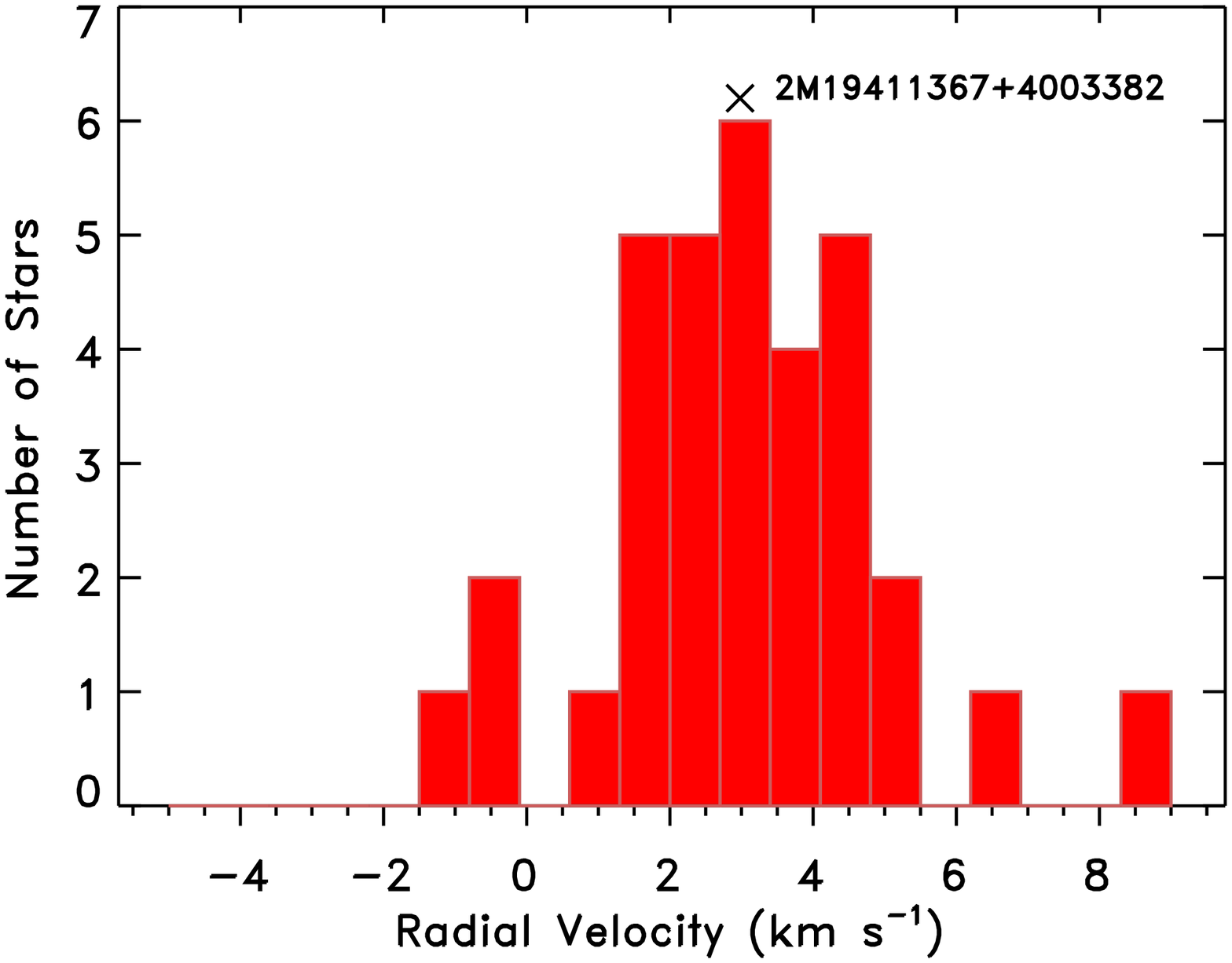} 
\caption{ Top: 2MASS color-magnitude diagram of \thecluster\ member stars (black squares),  \thestar\  (large  blue circle),  the three comparison stars (small orange  circles), and the star overlapping the optical study of \citet[][purple asterisk]{2001AJ....121..327B}. Bottom: Radial velocity histogram of \thecluster\ members. The $\times$ marks the RV of the Li-rich star.\label{fig:membership}}
\end{figure}

What makes \thestar\ such a powerful  test of Li-regeneration hypotheses is its apparent membership in an open cluster, which allows a direct comparison of its abundances to other stars  of nearly identical evolution.  However, its membership is not  conclusively established. Its radial velocity (RV), original proper motion measurement \citep{sanders72}, color, and magnitude (e.g, AT13) support membership (see Figure~\ref{fig:membership}), but its asteroseismic properties \citep{Stello:2011hu} and more recently measured proper motion \citep{Platais:2013im} support non-membership.  Both of these non-membership criteria have caveats. 
\cite{Platais:2013im}  reported that stars with high RV membership probabilities but low proper motion membership  probabilities tend to have larger proper motion errors that  stars of similar brightness. This suggests that there may have been confusion in the star's identification on the photographic plates. The asteroseismic properties of the Li-rich star suggest  a lower $\log g$  and mass than expected for cluster membership, but they yield a luminosity that places the star near the distance of the cluster. 
 (These caveats are discussed in more detail in  Section \ref{sec:case2}).

In this paper, we use   high-resolution, infrared spectra to address questions on the Li-rich star's cluster membership and evolutionary status.  We   select three similarly evolved comparison RGs in NGC 6819 (Section \ref{sec:comps}) and measure abundances of ten elements in the Li-rich star and comparison stars (Section \ref{sec:data}).  We measure a spectroscopic $\log g$, and find that it is consistent with the asteroseismic $\log g$  (Section \ref{sec:gravity}). The APOGEE spectra also reveal that the Li-rich star is rotating slightly more rapidly than the other RGs  (Section \ref{sec:vsini}). We combine our data and literature resources to consider possible contamination from companions or unrelated background sources (Section \ref{sec:binary}). Finally, we explore the implications of these results under the two possible cases of cluster membership (Section \ref{sec:overview}) and  summarize our findings (Section \ref{sec:end}).

 \begin{deluxetable*}{lcccc}
\tablecolumns{5}
\tablewidth{0pc}
\tabletypesize{\scriptsize}
\tablecaption{Measured Properties of NGC 6819 Red Giants \label{tab:stars}}
\tablehead{
   \colhead{Property} &
  \colhead{Li-rich} &
  \colhead{Comparison \#1} & 
  \colhead{Comparison \#2} & 
  \colhead{Comparison \#3}  \\  
 \colhead{} & \colhead{} &
 \colhead{(Most Similar)} & \colhead{(Near RC)} & \colhead{(Near  Bump)} }
\startdata
 2MASS  & 2M19411367+4003382  & 2M19404965+4014313 & 2M19412222+4016442 & 2M19412176+4012111 \\
KIC & 4937011 & 5111940 & 5112744 & 5112734 \\%
 $T_{\rm eff}$ (photometric) &  4700 & 4725 & 4700 &  4620  \\%
 $T_{\rm eff}$ (ASPCAP) & 4670 & 4687 & 4645 & 4632 \\%
 $\log L/L_{\rm \sun}$ & 1.52 & 1.59  & 1.67 & 1.69 \\
 $\log g$\tablenotemark{a}& 2.8 & 2.7 & 2.6 & 2.6 \\
$\xi$ (\kms) & 1.5 & 1.5  & 1.5 & 1.5 \\
\ali\ (dex)\tablenotemark{b}  & 2.3 & $\lesssim 0.7$ & $\lesssim 0.7$ & $\lesssim 0.7$\\%
\hline
\multicolumn{5}{c}{\emph{APOGEE-derived}} \\
 $\log g$  & 2.35 & 2.53 &  2.51 & 2.62 \\
 $A(^{12}$C) &  8.48$\pm$0.06 & 8.34$\pm$0.05 & 8.39$\pm$0.05 & 8.38$\pm$0.02 \\%
 $A(^{14}$N) & 8.11$\pm$0.05   &  8.15$\pm$0.02 & 8.13$\pm$0.02 & 8.16$\pm$0.03 \\
 $A(^{16}$O) &  8.94$\pm$0.1  & 8.82$\pm$0.05 & 8.75$\pm$0.04 & 8.77$\pm$0.02 \\%
 C/N & 2.3 & 1.6 & 1.8  & 1.7  \\
 \cratio & 25$\pm$5 & 24$\pm$5 & 25$\pm$5 & 20$\pm$8\\
$A$(Na) & 5.98$\pm$0.14 &  6.28$\pm$0.18  & 6.20$\pm$0.04 & 6.19$\pm$0.09 \\
$A$(Mg) &7.44$\pm$0.11 &  7.48$\pm$0.11   & 7.45$\pm$0.13 & 7.47$\pm$0.12 \\
$A$(Al) & 6.32$\pm$0.03   &  6.47$\pm$0.05  & 6.45$\pm$0.04 & 6.46$\pm$0.04 \\%
$A$(Si) & 7.50$\pm$0.06  & 7.51$\pm$0.06& 7.51$\pm$0.08 & 7.53$\pm$0.09 \\

$A$(Ti) & 4.93$\pm$0.17 & 5.02$\pm$0.07 & 5.02$\pm0.04$  & 5.03$\pm$0.12 \\%
$A$(Fe) & 7.45$\pm$0.12 & 7.52$\pm$0.04 & 7.48$\pm$0.12  & 7.48$\pm$0.08 \\
$A$(Ni) & 6.22$\pm$0.07 & 6.19$\pm$0.06 & 6.22$\pm$0.08 & 6.22$\pm$0.05 \\%
\vsini\ (\kms) & 8.5$\pm$1.1 & 6.6$\pm$1.6  & 4.4$\pm$1.5& 5.8$\pm$1.8 \\

 \hline
\multicolumn{5}{c}{\emph{APOKASC}} \\
$\nu_{\rm max}$  ($\mu$Hz) &  29.51 &  52.33 & 43.69 & 40.11\\%
$\Delta\nu$ ($\mu$Hz) & 4.15 &  5.16 & 4.39 & 4.12 \\
$\log g$ & 2.37  & 2.62 & 2.54 & 2.50 \\
$R/R_\sun$     & 9.1 & 10.4& 12.0& 12.4\\%
$M/M_\sun$     & 0.7& 1.7& 1.8& 1.8\\
$\log L/L_\sun$ &1.56 & 1.69& 1.80& 1.80
\enddata 
\tablenotetext{a}{$\log g$ from the isochrone and used to derive abundances.}
\tablenotetext{b}{Measurement for \thestar\ from AT13;  their Figure 3 provides estimates for comparison stars.}
\end{deluxetable*}
   \section{Selection of Comparison Stars}
\label{sec:comps}
The stars in this paper were observed with the Sloan Digital Sky Survey (SDSS) telescope \citep{2006AJ....131.2332G} and the Apache Point Observatory Galactic Evolution Experiment (APOGEE) instrument \citep{2008AN....329.1018A,Majewski:2012wx}  as part of the open cluster calibration sample for the APOGEE survey \citep{2013AJ....146...81Z,meszaros02}.
We analyzed the Li-rich star and three similarly evolved RGs in NGC 6819, using an infrared CMD to select  stars most similar to the Li-rich star, as illustrated  in Figure~\ref{fig:membership}.
All of the  \thecluster\ member stars observed by APOGEE are shown in Figure~\ref{fig:membership} together with  a 2.25~Gyr, $Z=0.023$ isochrone \citep{Marigo:2008fy} shifted by  $(m-M)_0=11.85$ \citep{Basu:2011cc}, and $E(J-K_s)=0.07$
 (from E($B-V$) = 0.14~mag, \citealt{2001AJ....121..327B}).
The Li-rich star sits below  the RC and blueward of the luminosity bump of the isochrone. The same is true of the most similar control star (2M19404965+4014313).
Another comparison star is near the red edge of the RC (2M19412222+4016442), and the remaining comparison star (M19412176+4012111) overlaps the luminosity bump within the uncertainties.
   The 2MASS and KIC identifiers of these fours stars are listed in Table~\ref{tab:stars}.  
Additionally,  2M19411476+4011008 was selected for analysis  solely because it has previously reported spectroscopic abundances and $\log g$ derived from optical spectra  \citep[][hereafter B01]{2001AJ....121..327B}. This star, identified as 978 in B01, is hotter  than the Li-rich star by $\sim200$~K and sits at the blue edge of the RC.  Because this star is much more evolved, it is not a good comparison star to the Li-rich star, and we provide results for this star separately in Table~\ref{tab:compA}.
  \begin{deluxetable}{lllc}
\tablecolumns{3}
\tablewidth{0pc}
\tabletypesize{\scriptsize}
\tablecaption{Properties of 2M19411476+4011008 compared to literature\label{tab:compA}}
\tablehead{
\colhead{Property} &  \colhead{this work}  &    \colhead{B01}&    \colhead{$\Delta$ (this work $-$ B01)} }
\startdata
 $T_{\rm eff}$  &  4920  &  4855 & +65 \\ 
 $\log L/L_{\rm \sun}$ &  1.79   &  &  \\ 
 $\log g$\tablenotemark{a} & 2.6 & 2.60 & +0.00 \\
$\log g$\tablenotemark{b}  & 2.58 & 2.60 &  $-0.02$\\
$\xi$ (\kms) & 1.5  & 1.26 & +0.24\\
 $A(^{12}$C) & 8.32$\pm$0.08 & \nodata & \nodata \\
 $A(^{14}$N) & 8.21$\pm$0.03 & \nodata & \nodata \\
 $A(^{16}$O) & 8.74$\pm$0.16 & \nodata & \nodata \\
 C/N & 1.3&  \nodata & \nodata \\
 \cratio & $>20$ & \nodata & \nodata \\
$A$(Na) & 6.46$\pm$0.07 & 6.96 & $-0.50$  \\
$A$(Mg) & 7.56$\pm$0.09& 7.65 & $-0.09$ \\
$A$(Al) &  6.56$\pm$0.02 & 6.44 & $+0.12$ \\
$A$(Si) &  7.60$\pm$0.03  & 7.82$\pm$0.14 & $-0.22$ \\         
$A$(Ti) &  5.04$\pm$0.08 & 5.22$\pm$0.16 & $-0.18$\\           
$A$(Fe) & 7.58$\pm$0.06 & 7.60$\pm$0.12  & $-0.02$\\           
$A$(Ni) & 6.28$\pm$0.06 & 6.37$\pm$0.14 & $-0.09$  \\        
\vsini\ (\kms) & 3.4$\pm$1.3 & \nodata & \nodata
\enddata
\tablenotetext{a}{$\log g$ from the isochrone.}
\tablenotetext{b}{$\log g$ from ionization balance.}
\end{deluxetable}

From the 2MASS colors and isochrones, we obtain the temperature, gravity and luminosity listed in the top section of Table~\ref{tab:stars}. 
\teff\ is the average of two $(J-K_s)_0$ to \teff\ calibrations (\citealt{1998A&A...333..231B} and \citealt{2009A&A...497..497G}).
The uncertainty in  $(J-K_s)_0$ from the 2MASS photometry is typically 0.028~mag, leading to a temperature uncertainty of $\sim80$~K in each calibration. 
The \cite{1998A&A...333..231B} calibration uses the derived color indices for ATLAS9 overshoot models (their Table 1) and gives slightly cooler temperatures than the \cite{2009A&A...497..497G} calibrations by 40--60~K.
This systematic difference in temperature is well within the 1$\sigma$ uncertainties of each calibration.
The resulting temperatures  are also in good agreement (all within 55~K) with those derived by the APOGEE Stellar Parameters and Chemical Abundance Pipeline (ASPCAP, Garc\'{i}a P\'{e}rez, et~al., in prep.). We can also tie this temperature scale to an independent spectroscopic analysis.  The comparison star overlapping the B01 study has $T_{\rm eff}=4855$~K derived from optical spectra compared to $T_{\rm ASPCAP}=4867$~K, and our $T_{\rm eff}=4920$~K. 

 Assuming a stellar mass of 1.7~\msun\ from the isochrone, we derive the surface gravities in the top section of Table~\ref{tab:stars}. These are the $\log g$'s  used to derive abundances. 
For all of the models, we adopted a microturbulence $\xi$ of 1.5~\kms, following the prescription adopted by ASPCAP ($\xi=2.24-0.3\log g$ )\footnote{See http://www.sdss3.org/dr10/irspec/aspcap.php\#aspcap} for $\log g=2.6$~dex.

\section{Abundances from APOGEE Spectra}
\label{sec:data}

 The  spectrum for each star has been processed with the APOGEE pipeline \citep{2015arXiv150103742N}. 
The observed data used here are the continuum-normalized ``aspcapStar''   spectra, with the exception of the wavelength region near the  \ion{Ti}{2} line, which falls near one of the detector gaps and is masked out in the ``aspcapStar'' spectra.  This spectral region comes from the ``apStar'' spectra, which we continuum normalize.  
All spectra are described and are  available in the tenth data release \citep[DR10,][]{2014ApJS..211...17A} of the SDSS-III \citep{2011AJ....142...72E}\footnote{https://www.sdss3.org/dr10/}.   

Abundances are  measured via spectrum synthesis using MOOG \citep{Sneden:1973el}\footnote{MOOG is available at http://www.as.utexas.edu/$\sim$chris/moog.html} 
and the APOGEE linelist (Shetrone et.~al., in prep.), focusing on small subsets of atomic and molecular features identified for each element by 
\citet[][hereafter S13]{2013ApJ...765...16S}.
All of the wavelengths in this paper refer to air wavelengths.
 A spectrum of a  $\sim$20~\AA\ region around each feature under consideration was generated using MOOG,  smoothed by a Gaussian to match the broadening of the observed spectra. 
Small adjustments were made to  the continuum level and velocity shift of the observed data to best match the models. 

\subsection{C, N, O and \cratio}
\label{sec:cno}
The abundances of C, N, and O  must be found iteratively because they are inter-related by  molecular equilibrium conditions.  We first measure C abundances from CO lines,  then fix the C abundances to measure O using OH lines. These steps are iterated until convergence is reached.  The N abundances can then be derived from CN lines. The lines used in this analysis are from Table 4 of S13, excluding  the (4--1)V-R $^{12}$C$^{16}$O,  (3--1)P$_2$9.5 $^{16}$OH, and  (0--1) R1 68.5 $^{12}$C$^{14}$N lines.
These abundances  and the C/N values are listed in the middle section of Table~\ref{tab:stars} for the Li-rich star and comparison stars and in Table~\ref{tab:compA} for  2M19411476+4011008. The uncertainties are the standard deviations of the individual measurements, and we address the uncertainties propagated from the stellar parameter uncertainties in the next section.
The three comparison stars all have nearly identical abundances, but the Li-rich star appears to have somewhat higher O.  The low C/N  confirms that standard FDU has completed and Li dilution should have occurred in all of the stars.  

\begin{figure}[tb]
\includegraphics[width=0.5\textwidth]{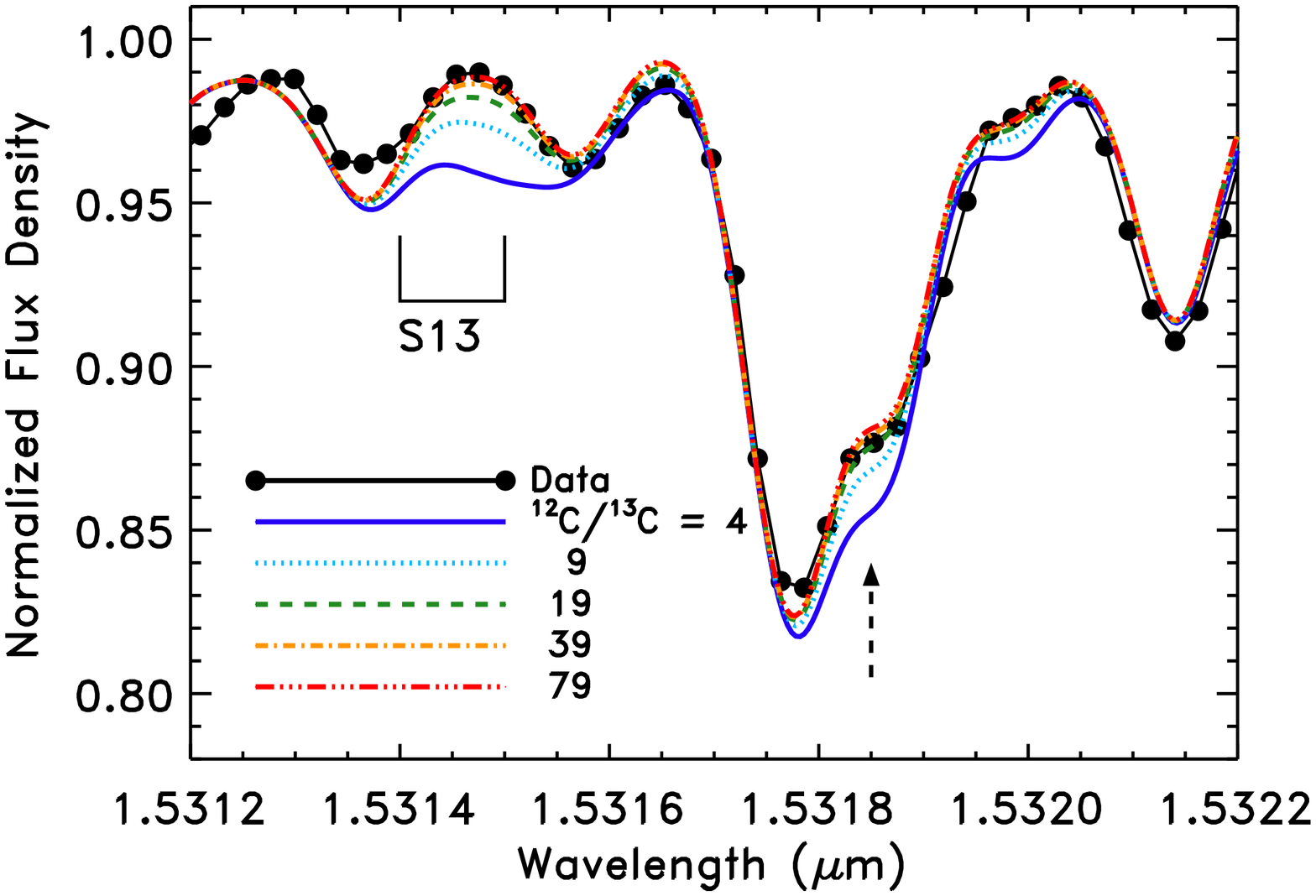} 
\includegraphics[width=0.5\textwidth]{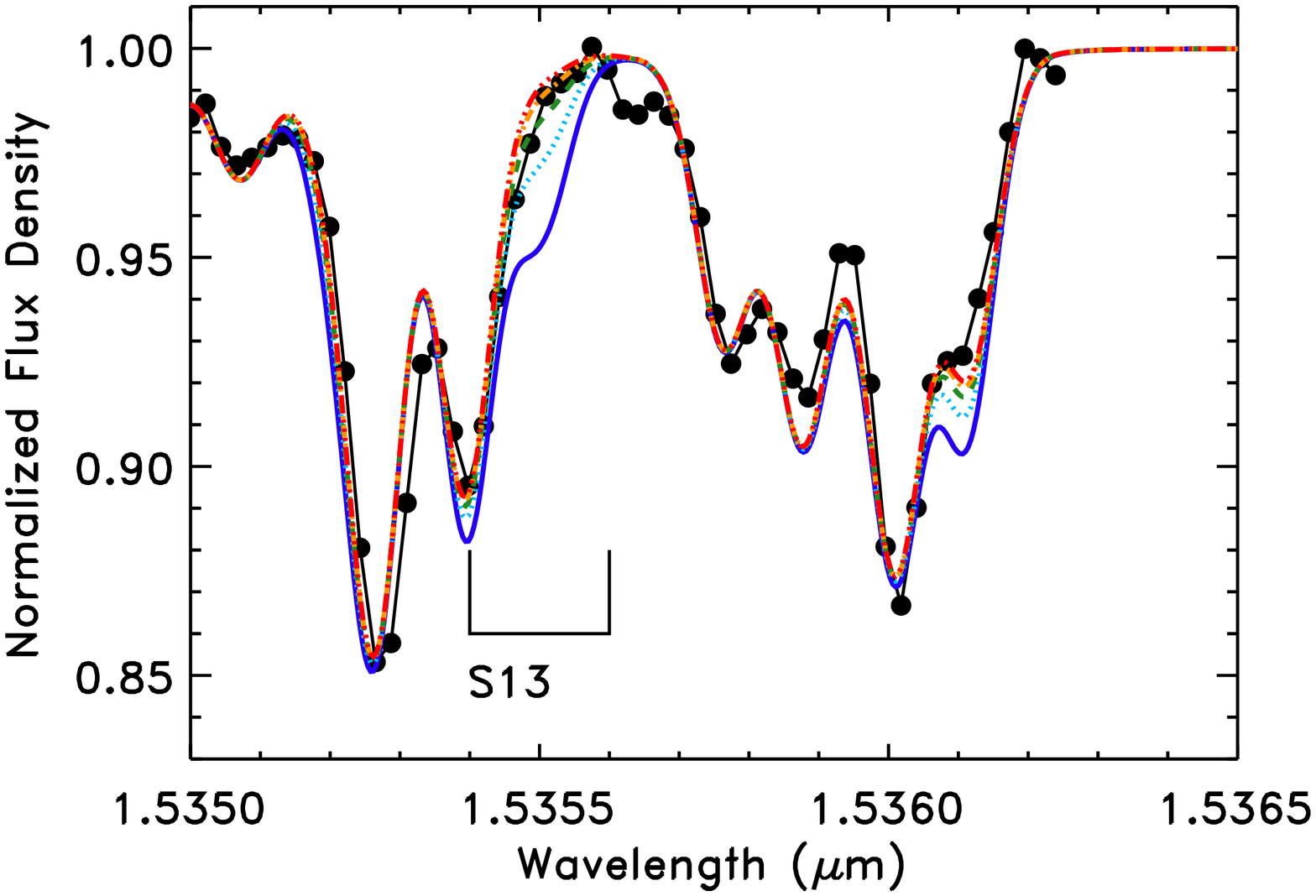} 
\caption{ Measurement of \cratio\ from  $^{13}$C$^{14}$N lines for the Li-rich star (circles). The wavelength intervals of S13 are marked. The synthetic spectra (colored lines) span a range of \cratio. 
Top: Low \cratio\ are ruled out in the S13 wavelength interval, but continuum level uncertainties make it difficult to place a stringent lower limit.  A neighboring region (arrow) suggests \cratio$\sim$20. Bottom:  The wavelength interval of S13 favors \cratio$\sim 30$.
 \label{fig:cratio}}
\end{figure}

To measure \cratio, we only used the  two $^{13}$C$^{14}$N lines in S13 because the $^{13}$C$^{16}$O features are weak and  contaminated with  night sky emission lines.  The  $^{13}$C$^{14}$N lines are also quite weak in all of the stars analyzed, which indicates high \cratio. We illustrate the measurement  of \cratio\ from the Li-rich star's spectrum  in Figure~\ref{fig:cratio}. Low \cratio\ ($\lesssim 15$) are clearly excluded by the weakness of the $^{13}$C$^{14}$N. The fits favor \cratio$\sim$20--30, consistent with normal levels of isotopic $^{13}$C enrichment in the stellar envelope following FDU.  The constraints against high \cratio, i.e., $>30$, are obviously weaker than those against low \cratio, but our C/N measurement gives us confidence that FDU has occurred. 

\begin{figure}[tb]
\includegraphics[width=0.5\textwidth]{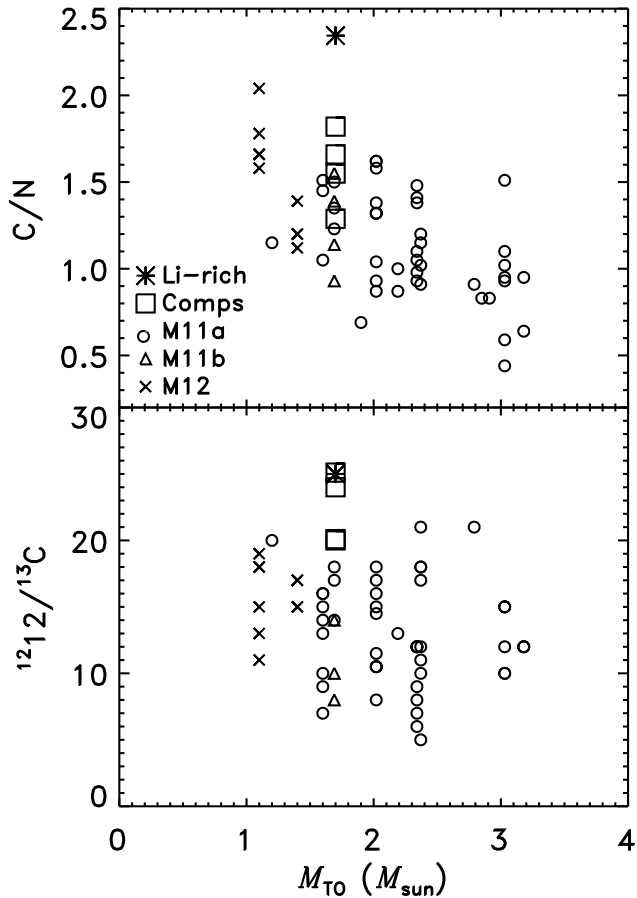} 
\caption{Relationship between cluster turn-off mass and C/N (top) and \cratio\ (bottom). The Li-rich star (asterisks) and comparison stars (squares) are compared to other open clusters in the literature (Mikolaitis et~al.\ 2011a, circles;  2011b, triangles;  2012, $\times$).
   \label{fig:cabun}}
 \end{figure}

In Figure~\ref{fig:cabun}, we compare our mixing indicators (C/N and \cratio) to red giants in other open clusters.  We plot the mixing indicators as a function of the clusters' turn off mass ($M_{\rm TO}$).  The literature values come from a series of papers focusing on the abundances of open cluster stars \citep{2011MNRAS.413.2199M, 2011MNRAS.416.1092M, 2012A&A...541A.137M}.  The first paper compiles even earlier studies from \cite{gilroy89, Luck:1994iu, 2000A&A...360..499T, 2005A&A...431..933T, Smiljanic:2009ik} and \cite{2010MNRAS.407.1866M}.  Figure~\ref{fig:cabun} demonstrates that the stars in this paper generally fit within the trends defined by the other clusters. The exception is  C/N for the Li-rich star, which is somewhat high for a star of its presumed mass. 
Oxygen is not processed in stars of this mass range, so we expect that [O/Fe] to remain near zero.  We find that the comparison stars have  [O/Fe] between $-0.03$ and $0.11$, which is consistent with our expectation given the uncertainties of $\sim 0.1$~dex. However, the Li-rich star again stands out. It has [O/Fe]$=0.3\pm0.16$~dex, suggestive of an unusual enhancement in O. Extra deep mixing would be expected to lower the [O/Fe], e.g., as in the case of IRS 7 \citep{2000ApJ...530..307C}.

\subsection{Na, Mg, Al, Si, Ti, Fe, and Ni}
\label{sec:metals}

The full line lists in Table 5 of S13 were used for  \ion{Fe}{1}, \ion{Mg}{1}, and \ion{Al}{1}. Subsets of the S13 list were used for  \ion{Si}{1} (15376.8, 15960.1, 16060.0, 16094.8, 16215.7, 16680.8,  16282.2~\AA),  \ion{Ni}{1}  (15632.7, 16584.4, 16589.3, 16673.7~\AA), and 
   \ion{Ti}{1}  (15543.756~\AA\  and 15602.842~\AA).
Sodium abundances were measured from two \ion{Na}{1} lines (16373.9 and 16388.9~\AA) identified in \cite{cunha14}. 
We fixed the C, N, and O  abundances found in Section \ref{sec:cno} for these analyses. 
The abundance results are given in Table~\ref{tab:stars} for the main sample, and the uncertainties are again the standard deviation of the measurements. 

In Table~\ref{tab:compA} we compare our results for 2M19411476+4011008 to the abundances derived from optical data by B01.  Columns two and three list the abundances and standard deviations ($\sigma$) measured here and in B01, respectively. When only one line for a given element was measured by B01, we do not list  $\sigma$.  The last column shows the differences between the abundances derived here and those derived by B01.  Most of our abundances agree within the quoted uncertainties. The exceptions are Si and Na. For both of these elements, we find lower abundances than B01.
Compared to  Comps 1--3, we generally measure slightly larger abundances for  2M19411476+4011008, although only Al disagrees with the other cluster stars outside the uncertainties.
\begin{figure}[t]
\includegraphics[width=.5\textwidth]{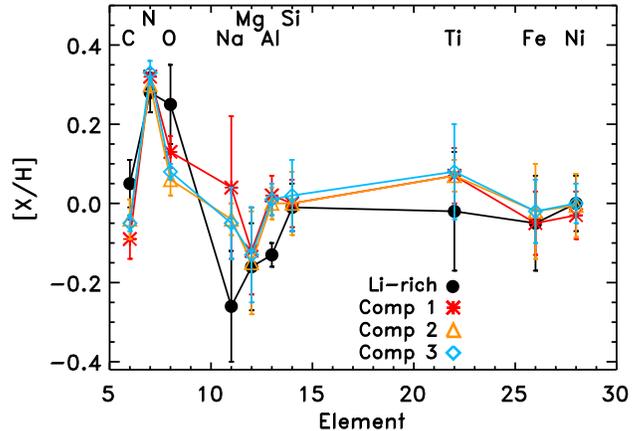} 
\caption{Elemental abundances  normalized to solar \citep{2009ARA&A..47..481A} of the Li-rich star and three comparison stars. \label{fig:abun}}
\end{figure}

In Figure~\ref{fig:abun}, we present our elemental abundance measurements (solar values  from \citealt{2009ARA&A..47..481A}) of the Li-rich star and the three comparison stars. Since all of these stars have similar
evolutionary stages and stellar parameters, we would expect them to all have similar abundances. (We exclude the star overlapping B01 from this plot because it is at a much more evolved evolutionary stage.)
The abundances of the Li-rich star are generally in good agreement with the Li-normal comparison stars, with the exceptions being O, Na, and Al.  
The enhanced $A$(O) was already mentioned in the previous section.    The $A$(Al) of the Li-rich star  is $\sim$0.14~dex lower than the three most similar comparison stars---well outside the 0.05~dex uncertainties.  $A$(Na) is also  quite low in the Li-rich star, though the uncertainties are large. However, this large uncertainty means that the two Na lines disagree. If we directly compare each line in the Li-rich star to the most similar control star, we find that both Na lines are significantly weaker in the Li-rich star, pointing to a true deficiency  in Na.   Thus, both $A$(Al) and $A$(Na) may be low for the Li-rich star.  If the reverse were true, the anomalous abundances could be explained by  extra-mixing, which enhances the surface abundance of Li, Na, and Al. 

\subsection{Abundance Sensitivities}
To estimate the sensitivity of the abundance measurements to uncertainties in the stellar  parameters used to select a model atmosphere for each star, we selected a  fiducial model of $T_{\rm eff}=4700$~K, $\log g=2.7$~dex, [M/H]=0.0, and $\xi=1.5$~\kms\ and tested  how the output abundances  change for perturbations of $\Delta T_{\rm eff}=\pm80$~K, $\Delta \log g=\pm0.2$~dex, $\Delta$[M/H]~$=\pm 0.1$~dex,  and $\Delta\xi=\pm 0.2$~\kms.  This was accomplished by adopting mean abundances of each element from our stellar sample, calculating the equivalent width needed to produce that abundance with the fiducial model for each line, and fixing the equivalent width to see how the output abundance changed with changes in the model atmosphere. The changes for all lines of a given element are then averaged and presented in Table~\ref{tab:deltaA}.  Each entry in the table gives two numbers. The first number is the change in abundance  when increasing the stellar parameter by the specified amount, and the second number is  the change in abundance  when the stellar parameter is decreased.   
For the sensitivities of C, N, and O, we adopted a simplified approach by using the line sensitivities of the three molecules present in our line list  as a proxy for either C, N, or O. 
The molecules in Table \ref{tab:deltaA} have their associated elements given in parentheses.

 \begin{deluxetable*}{lcccc}
\tablecolumns{5}
\tablewidth{0pc}
\tabletypesize{\scriptsize}
\tablecaption{Abundance sensitivities for the fiducial model \label{tab:deltaA}}
\tablehead{
\colhead{Species} &  
\colhead{$\Delta A/\Delta T_{\rm eff}$} &  
\colhead{$\Delta A/\Delta \log g$} &
\colhead{$\Delta A/\Delta $[M/H]} &
\colhead{ $\Delta A/\Delta \xi$} \\
\colhead{} &   \colhead{$\pm$80 K}  &    \colhead{$\pm$ 0.2 dex}  &  \colhead{$\pm$ 0.1 dex}  &   \colhead{$\pm$0.2 \kms}
 }
\startdata
CO  (C)&   +0.049/$-$0.055 &    +0.065/$-$0.063 &    +0.029/$-$0.042 &   $-$0.003/+0.002 \\
CN  (N) &   +0.037/$-$0.021 &    +0.053/$-$0.035 &    +0.017/$-$0.033 &   $-$0.014/+0.015 \\
OH  (O) &   +0.101/$-$0.097 &   $-$0.017/+0.040 &    +0.087/$-$0.107 &   $-$0.002/+0.002 \\
\ion{Na}{1} &    +0.042/$-$0.042 &   $-$0.007/+0.027 &   $-$0.002/$-$0.016 &   $-$0.002/+0.003 \\
\ion{Mg}{1} &    +0.042/$-$0.038 &   $-$0.041/+0.064 &    +0.018/$-$0.036 &   $-$0.012/+0.013 \\
\ion{Al}{1} &    +0.070/$-$0.071 &   $-$0.074/+0.102 &    +0.020/$-$0.044 &   $-$0.033/+0.030 \\
\ion{Si}{1} &    +0.021/$-$0.015 &   $-$0.044/+0.062 &    +0.040/$-$0.053 &   $-$0.024/+0.023 \\
\ion{Ti}{1} &    +0.111/$-$0.109 &   $-$0.001/+0.016 &   $-$0.005/$-$0.012 &   $-$0.045/+0.054 \\
\ion{Ti}{2} &   $-$0.019/+0.022 &    +0.078/$-$0.092 &    +0.035/$-$0.036 &   $-$0.035/+0.038 \\
\ion{Fe}{1} &    +0.050/$-$0.045 &   $-$0.032/+0.046 &   +0.026/$-$0.042 &   $-$0.040/+0.043 \\
\ion{Ni}{1} &    +0.019/$-$0.014 &    +0.006/+0.008 &    +0.020/$-$0.033 &   $-$0.017/+0.018
\enddata
\end{deluxetable*}

\section{Surface Gravity}
\label{sec:gravity}
\subsection{Spectroscopic  $\log g$}
\label{sec:logg}
\begin{figure}[tb]
\includegraphics[width=0.5\textwidth]{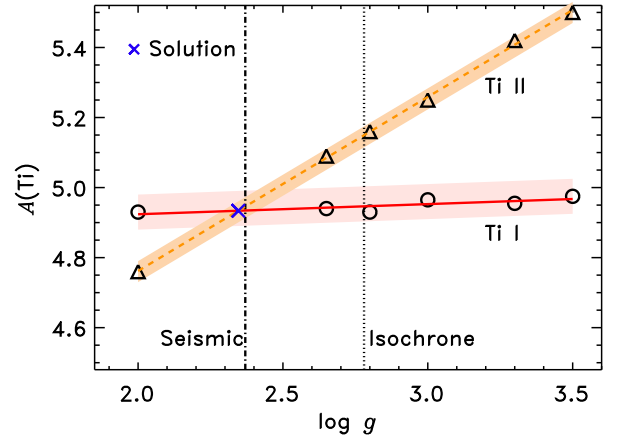}
\caption{Ionization equilibrium plot for the Li-rich star. The shaded regions around   \ion{Ti}{1} (circles) and \ion{Ti}{2}(triangles)  show individual line uncertainty 
  measurements  of  $\pm 0.05$ and $\pm0.03$~dex, respectively. The intersection of these relations ($\times$) yields the  $\log g$.  Vertical lines mark the expected $\log g$ from isochrones (dotted line) and asteroseismology (dash-dotted line). 
   \label{fig:ti}}
 \end{figure}

 For elements with low ionization potential that are predominantly ionized in the photospheres of the
    stars of interest, the strengths of lines of singy ionized atomic lines are  sensitive to the $\log g$ of the star, while neutral atomic lines are not.
When lines of both species of an element are available, they can be used to measure $\log g$. 
In the APOGEE spectral range, the \ion{Ti}{2} line at 15873.834~\AA\ is the sole known ionized line. \cite{2014ApJ...787L..16W} reported an experimentally measured $\log gf = -1.90\pm0.08$ for this line, while the APOGEE team adopted an astrophysical $\log gf=-2.06$ from fitting spectra of the Sun and Arcturus. 
The latter  is adopted here.
 We measured the strength of this \ion{Ti}{2} feature and  the \ion{Ti}{1} lines for our five stars while varying the $\log g$ of the atmosphere models from  2.0 to 3.5~dex, as illustrated in Figure~\ref{fig:ti} for the Li-rich star.
   Our spectroscopic   $\log g$  is the one for which  $A$(Ti) of   \ion{Ti}{1} and \ion{Ti}{2}  are equal.  These $\log g$ measurements are listed in the second section of Table~\ref{tab:stars}, and they generally agree well with those predicted from the cluster isochrone. The exception is the Li-rich star.   We find $\log g=2.35$~dex compared to the isochrone value of 2.8~dex; however, this is a good match to the $\log g$ derived from asteroseismology (Section \ref{seismo}).
   
  To check  the impact of having a single  ionized line with an uncertain oscillator strength, we check for a systematic offset in $\log g$ using the  star that overlaps the B01  study.  
    They measured $\log g$ with the ionization balance of Fe and find $\log g=2.6$~dex.   This  agrees with both our ionization balance  (2.58~dex) and  isochrone (2.6~dex) results.  

\subsection{Asteroseismic  $\log g$}
\label{seismo}

An additional gravity measure comes from the asteroseismic $\nu_{\rm max}$ in the APOKASC catalog \citep{pinsonneault14}, using the standard scaling relationship where $g \propto \nu_{\rm max}\sqrt T_{\rm eff}$ \citep{1995A&A...293...87K}. These gravities are listed in the bottom section of Table~\ref{tab:stars}.  In all cases, the  asteroseismic gravities agree well with our spectroscopic $\log g$. 
In turn, the spectroscopic $\log g$ matches well with the isochrone measurement for all but the Li-rich star.
  The Li-rich star has a spectroscopic $\log g$ that is in good agreement with the asteroseismic value, but it is  0.4~dex {\it lower} than what is expected from isochrones.
Unfortunately, 2M19411476+4011008  does not have a $\nu_{\rm max}$  measurement  so we do not have a direct comparison between the asteroseismic $\log g$ and $\log g$ derived from ionization balance of Fe lines.

\section{Rotational velocity}
\label{sec:vsini}
While measuring abundances, we required a slightly larger broadening of the synthetic spectra to fit the lines of the Li-rich spectrum compared to the other spectra. We acquired the  rotational velocities of the stars in this study from the pipeline being developed for APOGEE (Bizyaev et~al., in prep.). Briefly, the spectra are cross-correlated with synthetic templates that have been convolved with a rotational broadening profile for a range of \vsini.
The results of this analysis are listed in Table~\ref{tab:stars}. As expected,  the Li-rich star has the largest \vsini, and the comparison stars have \vsini\ near the detection limit for APOGEE's resolution. The clear measurement of \vsini$=8.5$~\kms\ points to unusually fast rotation in the Li-rich star.  Such a high \vsini\ for RGs is rare. In a recent survey for fast rotators among open cluster RGs, \cite{2014AJ....147..138C} found only one RG  out of $\sim 270$ cluster members that rotate as rapidly as the Li-rich star.

\section{Binary Companions or Other Blends}
\label{sec:binary}

We performed multiple checks of whether our spectroscopic analysis is  affected by either a companion to the Li-rich star or an unrelated fore/background object. 
First, we fit model spectral energy distributions (SED) to photometry spanning optical  \citep{TabethaHole:2009gi}, near-IR  \citep[2MASS, ][]{2006AJ....131.1163S},   and mid-IR  \citep[WISE, ][]{2010AJ....140.1868W} wavelengths, allowing the temperature of the models and visual extinction to be free parameters.  The best results are obtained for a bare photosphere  with a  hotter temperature (\teff$\sim5000$~K) than that derived from our $(J-K_s)_0$ calibration and a larger line-of-sight extinction.  This solution implies that the Li-rich star is a more distant star affected by additional extinction. However, the analysis favors the largest extinction we allowed in the fit, $A_V=0.665$, which is slightly higher than  $A_V=0.623$, the extinction in the \cite{1998ApJ...500..525S} map at this position. 
Additionally, the same SED analysis of the most similar comparison star also favors a  hotter, more reddened solution. Therefore, we conclude that the temperature difference may  be  systematic and cannot discriminate the Li-rich stars's cluster membership.  We also find no evidence for an IR excess out to 10~$\mu$m.

APOGEE made three unique observations of the Li-rich star,  roughly evenly spaced over the course of a month. 
The RV is constant within the 0.1--0.2~\kms\ uncertainties of the three observations, and the Li-rich star's RV matches that of the cluster, as illustrated in Figure~\ref{fig:membership}. Admittedly, one  month is a short time span to look for RV variations, but  the literature contains additional RV measurements for this star spanning at least 20 years. The optical spectrum taken by AT13 was taken between 2010--2012 and also had an RV consistent with cluster membership.
 \cite{TabethaHole:2009gi} define the star as a single, cluster member based on four RV measurements going back at least to 1992.
Taken together, all of these RVs suggest that \thestar\ is  a RV stable cluster member.

We  also leveraged the individual visit spectra to search for evidence of a secondary spectrum by  cross-correlating each   visit spectra with a library of template spectra. All three  give best fits to templates of $T_{\rm eff} \sim 5000$~K. The  shape of the cross-correlations peaks do not change between visits, and no  secondary peaks are found, ruling out bright companions of similar spectral type.
A second round cross-correlations used the TODCOR algorithm \citep{1994ApJ...420..806Z} to generate a two-dimensional cross-correlation using a wide range of possible secondary spectral types: no significant secondary peaks were identified. 
We  conclude that the spectral absorption features are truly arising in the red giant's atmosphere and are not significantly  contaminated by flux from any binary companion or unassociated blend.

\section{Interpretation of the Li-rich Star}
\label{sec:overview}
Our analysis of the Li-rich star has confirmed that it is evolved---its low $\log g$ and non-solar C/N are consistent with the picture of a RG that has undergone FDU and should have depleted Li. However, given the uncertainties in the cluster membership, the Li-rich star may be a more massive, more luminous background RG that coincidentally shares the same RV, [Fe/H], and CMD locus as the cluster.  Therefore, to fully explore the implications of our measurements we provide two scenarios. The first scenario  presumes that the  Li-rich star is not a member of \thecluster, and we explore how that assumption changes our derived abundances, $\log g$, etc. The second scenario explores the implications of the star being a cluster member, and we demonstrate why we favor this scenario.

\subsection{Case 1: Cluster Non-Membership}
\label{sec:case1}
\subsubsection{Underestimated Reddening}
  If the Li-rich star is not a member of the cluster, then the temperature derived from the $J-K_s$ is a lower limit, since the reddening may be underestimated.  
We can test how our analysis will change if we adopt the total line-of-sight reddening observed in the direction of \thestar.  
 If we adopt $A_V=0.665$ (the preferred solution of the SED fitting), then $E(B-V)=0.215$ and $E(J-K_s)=0.107$. Reapplying our temperature calibration yields 4840~K.   
 The strong temperature dependence of \ion{Ti}{1}  lines increases $A$(\ion{Ti}{1}), while  \ion{Ti}{2} remains nearly constant.  Thus, the ionization equilibrium plot (Figure~\ref{fig:ti}) for $T_{\rm eff}=4840$~K 
 would have a similar shape, but the horizontal line would shift upwards by $\sim0.2$~dex and intersect the  $A$(\ion{Ti}{2}) line at  $\log g=2.81$ and $A$(Ti)$=5.14$~dex.
This means that if the Li-rich star is in indeed a background red giant, then its surface gravity is higher than that predicted by the asteroseismology and matches that expected from the stellar isochrones.

Since the Li-rich star has the same metallicity as the  confirmed cluster stars, we can adopt solar metallicity stellar evolution tracks  \citep{Bertelli:2008ge,Bertelli:2009ic} to find the range of masses and ages of stars near this $T_{\rm eff}$ and $\log g$ but that have luminosities exceeding that adopted in Table~\ref{tab:stars} under the assumption of cluster membership. We find that stars within $T_{\rm eff}=4840\pm80$~K, $\log g=2.81\pm0.2$~dex, and $\log L/L_\sun \ge 1.52$ span a mass range of  1.8--3.0~\msun\ and an age range of 0.4--1.6~Gyr. The lowest mass stars in this range  must be RC stars rather than first ascent stars in order to meet all of the selection criteria. First ascent stars are too faint.   At higher masses,  both first ascent stars and core He burning stars  satisfy all three constraints.  For the entire mass range, the stars should have completed first dilution \citep[see, e.g., Fig.~1 of ][]{Charbonnel:2000ud} meaning the high Li abundance is still unexplained by standard stellar evolution.
 
 Using the sensitivities calculated in Section \ref{sec:metals}, we can conclude that increasing the temperature by 140~K has the following effects: Fe changes by +0.09,  C by +0.09, N by +0.06, O by +0.18, Na by +0.07, Mg by +0.07, Al by +0.12, Si by +0.04, Ti  (from ionization balance) by +0.16, and   Ni by +0.03.  
The new C/N is 2.5, which is still consistent with the completion of FDU.

 \subsubsection{Internal Li Replenishment}
   We can use mixing signatures to explore  whether  internal replenishment of Li is a viable mechanism for the full range of possible masses and ages of the Li-rich star. Constraints on these stellar parameters were made in the last section for the high extinction case, but we must also consider  the case where  the Li-rich star is affected by the same amount of extinction as the cluster stars, i.e., if  there is negligible extinction between the cluster and the Li-rich star. We repeat the analysis with stellar evolution tracks for $T_{\rm eff}=4700\pm80$~K, $\log g=2.35\pm0.2$~dex, and $\log L/L_\sun \ge 1.52$. The resulting mass range is $M=1.8$--4.0~\msun\ and an age range of 0.2--1.6~Gyr.  In this scenario, both first and second ascent evolutionary stages for the full mass range are consistent with the selection criteria. As a result, 1.8~\msun\ stars  near the cool end of our $T_{\rm eff}$ range (4620~K) are luminosity bump stars that also meet the gravity and luminosity constraint.
 
In either assumed extinction scenario, the Li-rich star is an evolved giant that should have undergone convective dilution and show depleted Li.
 The allowed parameters of the Li-rich star includes two regimes where Li regeneration has been suggested to occur---the luminosity bump, and the narrow mass range of RC stars identified by \cite{Kumar:2011jr}.  Thus, internal Li regeneration seems to be a viable explanation. 
However, a common feature of internal Li regeneration models is that the Li synthesis  must happen deep enough in the star that whatever mechanism is ferrying the newly synthesized Li into the convection zone must also ferry $^{13}$C and thereby lower the surface \cratio.
The absence of strong $^{13}$C features excludes the deep mixing that should accompany Li-regeneration via the D12 models favored by AT13, unless we are observing the star near the beginning of the rapid Li enrichment  stage (in which the star spends less than a few Myr with enriched Li but FDU levels of \cratio).  

 \subsubsection{External Li Replenishment}
 \label{sec:external}
External replenishment models run into difficulties with the amount of material required to raise the \ali\ to the observed value. By rearranging the equation for predicting  enhanced Li from planet engulfment in \citet[][their Eq.~2]{carlberg12}
 we can write  
 \begin{equation}
 \label{eq1}
 q_{\rm e}=\frac{10^{A({\rm Li})_{\rm new}} - 10^{A({\rm Li})_\star}}{10^{A({\rm Li})_{\rm p}}-10^{A({\rm Li})_{\rm new}}},
 \end{equation} 
  where $q_{\rm e}$ is the ratio of the mass of the polluting material to the mass in the stellar convective envelope, and the \ali\ subscripts of ``new,'' $\star$, and ``p'' refer to the post-engulfment, the normal stellar, and pollutant abundances, respectively.   Adopting \ali $_\star=1.49$~dex (the largest non-enriched Li), \ali $_{\rm new}=2.3$~dex (the observed abundance), and \ali $_{\rm p}=3.3$~dex, (the solar nebular value) we find that $q_{\rm e}=0.09$.
 We used the MESA stellar evolution code \citep[][version 6794]{2013ApJS..208....4P} to generate limiting mass models of $M=1.7$ and 4.0~\msun\ to determine the amount of mass in the outer convection zone for stars near 4700 and 4840~K at different phases of the RGB evolution.  In all cases, that mass was at least 1.1~\msun.  Consequently, the amount of material required to explain the enhanced Li in \thestar\  is 0.1~\msun, which is a stellar mass companion. 
 Such a low mass companion would have been missed by our binary analysis. However, low mass stars are also Li depleted. 
Substellar companions retain their birth supply of Li, but they are too low mass unless they are significantly enriched in Li over the solar nebular value.

\subsection{Case 2: Cluster Membership}
\label{sec:case2}
The inferences on the origin of the excess Li mentioned in the previous section are valid in the case of cluster membership, but we place further constraints on the unusual nature of the Li-rich star if it is truly a cluster member.  The cluster membership is supported from a number of lines of evidence.  Its RV is  consistent with membership, and its optical and infrared magnitudes coincide with the cluster's CMD locus. This latter piece of evidence is further supported by the fact that the temperature we derived from the 2MASS colors is the same as that derived by the ASPCAP pipeline (see Section \ref{sec:comps}). The ASPCAP pipeline fits model spectra to the continuum-normalized observed spectrum and is, importantly, independent of any assumption of reddening. Therefore, the ASPCAP temperature verifies that the reddening of the Li-rich star is the same as the rest of the cluster and thus  shares the temperature and magnitude of the cluster's isochrone.  %

 \subsubsection{A Deeper Look at Proper Motions}
\begin{figure}[tb]
\includegraphics[width=0.5\textwidth]{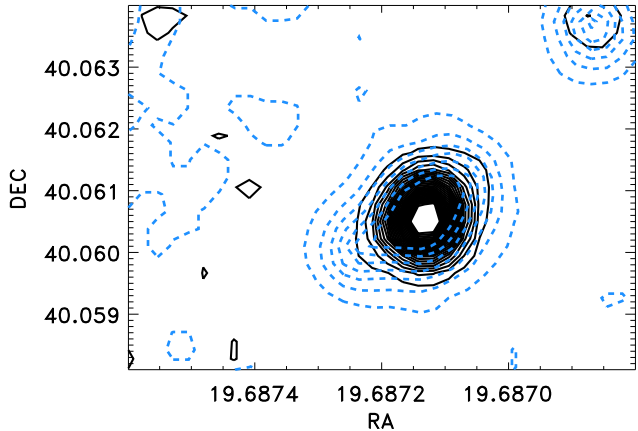} 
\caption{Image of the Li-rich star in 2MASS $H$-band image (solid lines) and the DSS optical blue image (dashed lines).  The optical image of the Li-rich star is slightly elongated on one side compared to the infrared image.
   \label{fig:image}}
 \end{figure}

One of the major pieces of evidence against membership for the Li-rich star is that \cite{Platais:2013im} find proper motions inconsistent with membership. However,  the authors of that paper noted that finding low proper motion membership probabilities for stars with high RV membership probabilities is likely due to source confusion (see their Figure 6).  In fact, an earlier proper motion  study \citep{sanders72}  gave the Li-rich star (star 90 in that paper) a 90\% membership probability. An inspection of a Digitized Sky Surveys (DSS) optical blue image and the 2MASS $H$-band image in Figure~\ref{fig:image} reveals an elongation   in the optical DSS image that is not present in the $H$-band image. This is suggestive of a faint, blue source.   
If this source is visible in the plates used by  \cite{Platais:2013im}, or worse is only visible in some of the plates, it could shift the apparent photocenter of the Li-rich star and  skew the proper motion measurement.    Star 90  on the \cite{sanders72}  plate is very round, suggesting that the nearby blue source is too faint to adversely affect the earlier proper motion measurement. Indeed, the $V$ limit of the \cite{sanders72} plate is quoted as $\sim 14.5$~mag, whereas the magnitude limits of the plates used in \cite{Platais:2013im}  are $V=17$--22~mag.
 Because this visual companion is faint and blue, it will not contribute significantly to  the H-band spectrum of the Li-rich star, consistent with our  analyses in Section \ref{sec:binary}. 

 \subsubsection{Problems with Replenishment Mechanisms}
Cluster membership adds more problems to both the internal and external replenishment models.  As a cluster member, the evolutionary stage of the Li-rich  star is clearly below the luminosity bump,  which led A13 to favor the recent D12 models.  However, there is a discrepancy between the D12 models  and the Li-rich star's CMD position.   D12's model predicts an evolution  that traces an  extended luminosity bump-like zigzag on the CMD. 
Unlike the luminosity bump,  the  extended zigzag  in the D12 models  parallels the first ascent RGB at {\it lower} luminosity. For a given $T_{\rm eff}$, the $\Delta \log L$ is $\approx -0.2$,  corresponding 
 to a magnitude difference of +0.5.  The Li-rich star does not sit 0.5 magnitudes below the RGB in either optical or infrared CMDs.
For external replenishment models, the other cluster RGs provide a revised estimate for pre-engulfment \ali, which is lower than that estimated in Section \ref{sec:external}. Additionally, with a well known evolutionary stage, we can more accurately pinpoint a MESA model to estimate the size of the convection zone. Using the 1.7~\msun\ MESA run, we inspected the internal profiles for four evolutionary stages on the first ascent RGB near $T=4700$~K. In all cases, there is $\sim 1.4$~\msun\ in the convective envelope, requiring 0.15~\msun\ of accreted material to explain the observed \ali.

 \subsubsection{An Unusually Low Mass for the Li-rich Star}
Another  implication of the cluster membership is that it allows us to put stronger  credence on  the asteroseismology result, since our spectroscopic $\log g$ agrees with the asteroseismic $\log g$.
The scaling relations for $\nu_{\rm max}$ and $\Delta\nu$ can both be cast to depend  only on \teff, $M$, and either $R$ or  $L$. Since \teff\  is accurately known and has the weakest influence on $\nu_{\rm max}$ and $\Delta\nu$, the remaining  quantities can be derived without any other assumptions.   These results are given in the bottom section of Table~\ref{tab:stars}.   There is good agreement between the luminosity derived through asteroseismology and that derived assuming the star is in the cluster. This analysis places  the Li-rich star at roughly the distance of the cluster but with a mass of only 0.7~\msun, which explains the ``low'' $\log g$.   The main sequence lifetime of such a low mass star exceeds the age of the universe. 
If this mass is correct, the  most plausible  explanation is that the star lost a significant fraction of its mass, and we suggest that this mass loss is related to the source of Li.
Furthermore, this low mass does not nullify the CMD position as evidence of cluster membership. The luminosity of stars with H-burning shells is dictated by the mass of the stellar core, leading to a core mass-luminosity relationship. \cite{1988ApJ...328..641B} demonstrated that this relationship holds for intermediate mass stars and that the total mass has a negligible effect on this relationship.  Thus, if the Li-rich star began its life with the same mass as the other RGs and only lost mass recently, it should have the same core mass as the other RGs.

 \subsubsection{External Enrichment at Low Mass}

The case for enrichment by an external source of \ali\ is much more feasible if the star is really 0.7~\msun. We can again adopt  \ali\ $<0.7$~dex for the pre-engulfment abundance and use the MESA models of the 1.7~\msun\ star.
If an engulfed companion strips 1~\msun\ of material from the star before it disperses  into the remaining stellar envelope, then the remaining envelope is only 0.4~\msun. Since the star is not entirely stripped of the original convective envelope, the mass loss not does extend deep enough to expose material with low \cratio. Only a 45$M_{\rm Jup}$ object is required to enrich \ali\ up to the observed value.  This mass is safely in the substellar companion regime of objects, which retain their  birth  \ali.  

 \subsubsection{Updated Abundances at Low $\log g$}
\begin{figure}[t]
\includegraphics[width=.5\textwidth]{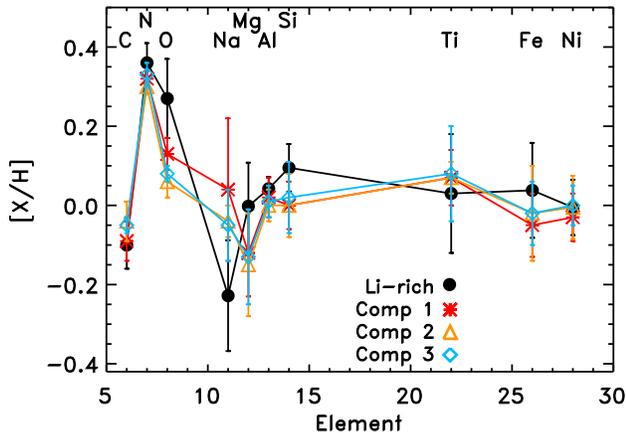} 
\caption{Elemental abundances  normalized to solar \citep{2009ARA&A..47..481A}.  The Li-rich star's abundances are computed for a $T_{\rm eff}$=4700~K and $\log g$=2.35~dex atmosphere model. The  abundances of the control stars are the same as in Figure~\ref{fig:abun}.  \label{fig:abun2}}
\end{figure}

   If the \thestar's $\log g$ is truly so low, then we must re-derive the abundances.  Because using the abundance sensitivities in Table~\ref{tab:deltaA} requires extrapolating by 0.2 dex beyond the $\log g$ range we tested, we instead calculate all of the  new abundances with the lower $\log g$ atmosphere.  We  calculate the equivalent widths needed to create the abundances we measured for the Li-rich star with the $\log g=2.80$~dex model and then find how the abundances changes for these same equivalent widths with a $\log g=2.35$~dex model. For the C, N, and O abundances from the molecular lines, we iterate the process until the output abundances match the  input abundances. The lower $\log g$ has the following effects:  Fe changes by +0.09, C by $-0.15$, N by $+0.08$, O by +0.02, Na by +0.03, Mg by +0.16, Al by +0.17, Si by +0.11, Ti  by +0.05, and Ni by $-0.005$~dex.   Figure~\ref{fig:abun2} shows these new abundances compared to that of the comparison stars. The updated $A$(Al)  is now in excellent agreement with the other cluster stars. The rather large increases in the abundances of Mg, Si, and Fe are still consistent with the abundances of the other cluster stars within the uncertainties. Furthermore, the C/N ratio is now 1.4, which is also in much better agreement with  other RGs in both \thecluster\ and in other clusters (Figure~\ref{fig:cabun}). However, oxygen is still unusually high, with [O/Fe]$\sim0.23\pm0.15$~dex.
   
 \subsubsection{Comparison to Other Known Li-rich Cluster Giants}
Finally, if \thestar\ is truly a member of \thecluster, then we can compare this star to the few other known Li-rich red giants  in other open clusters.  These include Trumpler 5 (Tr~5) \#3416 \citep{2014A&A...564L...6M} and  Berkeley 21 (Be~21) T33 \citep{1999A&A...348L..21H}. Both of these open clusters have significantly sub solar metallicities,  near $-0.5$~dex. In both cases, the Li-rich stars are super Li-rich, \ali $>3.3$~dex. Tr~5~3416 is also known to be a slow rotator and have \cratio=14.
The Li-rich star in Tr~5 sits on the RC, while the star in Be~21 sits a magnitude above the RC.  
Therefore, \thestar\ does not share many common properties with these other known Li-rich stars in open clusters. It is solar metallicity, fainter than its host cluster's RC, not super Li-rich, and has faster rotation and higher \cratio\ than the Tr~5 Li-rich star. Additionally, \thestar's initial mass was presumably 1.7~\msun, and \cite{2014A&A...564L...6M} estimates that the masses of the Tr~5 RC stars are 1.0--1.4~\msun, while \cite{2006AJ....131.1544B} gives 1.4~\msun\ as the mass of the turn-off stars in Be~21.

\section{Conclusions}
\label{sec:end}
We have performed a spectroscopic analysis  of the Li-rich star \thestar, and we have provided  crucial measurements of $\log g$, C/N, and \cratio\ that  confirm that the star is an evolved red giant with evidence of having completed  FDU  (in its C/N and \cratio) but showing no evidence of low \cratio\ that should accompany the previously favored model of explaining the star's enriched Li.
We have also measured abundances of eight additional elements and found a chemical similarity between the Li-rich star and the other cluster members  in elements that are not affected by nuclear processing, including both iron group (Fe, Ni) and $\alpha$ elements (Mg, Si, Ti). 

The membership of \thestar\ has recently been questioned by both asteroseismic and recent proper motion measurements. We have argued that an older proper motion measurement may be more reliable, and this older measurement is compatible with membership.
The asteroseismic non-membership evidence actually contains a surprising result.  The asteroseismic $\nu_{\rm max}$ yields a $\log g$ lower than that expected for members, which we confirm spectroscopically. 
Recomputing the abundances at lower $\log g$ brings the abundances of the Li-rich star in better agreement with the other red giants in \thecluster.
Furthermore,  $\nu_{\rm max}$ and $\Delta\nu$ also gives the star's luminosity and mass and places the Li-rich star at roughly the distance to the cluster, but with a mass of only $0.7$~\msun. Such a low mass star could not have evolved to the RG stage, implying that the Li-rich star has lost significant mass. The similar luminosity to other \thecluster\ RGs implies that the Li-rich star has a similar core mass to the other RGs, which is also consistent with the picture of extensive mass loss.
 For the case of external enrichment, this low stellar mass is the only case where a putative engulfed companion can have low enough mass to have retained its birth Li but still be massive enough to contribute enough Li to explain the observed stellar \ali. 
A more detailed analysis of the stellar oscillation spectrum may provide new insights on whether the scaling relationships do not apply to this star or whether the internal structure of the Li-rich star can shed light on its anomalous Li abundance and mass.

\acknowledgments
We thank the anonymous referee for providing valuable feedback on our manuscript.

JKC acknowledges partial support for this work by an appointment to the NASA Postdoctoral Program at the Goddard Space Flight Center, administered by Oak Ridge Associated Universities through a contract with NASA.  VS acknowledges partial support for
this research from the National Science Foundation (AST1109888).

Funding for SDSS-III has been provided by the Alfred P. Sloan Foundation, the Participating Institutions, the National Science Foundation, and the U.S. Department of Energy Office of Science. The SDSS-III web site is http://www.sdss3.org/.
SDSS-III is managed by the Astrophysical Research Consortium for the Participating Institutions of the SDSS-III Collaboration including the University of Arizona, the Brazilian Participation Group, Brookhaven National Laboratory, Carnegie Mellon University, University of Florida, the French Participation Group, the German Participation Group, Harvard University, the Instituto de Astrofisica de Canarias, the Michigan State/Notre Dame/JINA Participation Group, Johns Hopkins University, Lawrence Berkeley National Laboratory, Max Planck Institute for Astrophysics, Max Planck Institute for Extraterrestrial Physics, New Mexico State University, New York University, Ohio State University, Pennsylvania State University, University of Portsmouth, Princeton University, the Spanish Participation Group, University of Tokyo, University of Utah, Vanderbilt University, University of Virginia, University of Washington, and Yale University. 

This publication  uses  data products  both from the 
Wide-field Infrared Survey Explorer, which is a joint project of the
University of California, Los Angeles, and the Jet Propulsion
Laboratory California Institute of Technology, funded by the
National Aeronautics and Space Administration, and from the
Two Micron All Sky Survey, which is a joint project of the University of Massachusetts and the Infrared Processing and Analysis Center/California Institute of Technology, funded by the National Aeronautics and Space Administration and the National Science Foundation.

The Digitized Sky Surveys were produced at the Space Telescope Science Institute under U.S. Government grant NAG W-2166. The images of these surveys are based on photographic data obtained using the Oschin Schmidt Telescope on Palomar Mountain and the UK Schmidt Telescope. The plates were processed into the present compressed digital form with the permission of these institutions.


\begin{thebibliography}{49}

\bibitem[{Ahn {et~al.}(2014)}]{2014ApJS..211...17A} Ahn, C.~P., Alexandroff, R., Allende Prieto, C., et al.\ 2014, \apjs, 211, 17 

\bibitem[Allende Prieto et al.(2008)]{2008AN....329.1018A} Allende Prieto, C., Majewski, S.~R., Schiavon, R., et al.\ 2008, Astronomische Nachrichten, 329, 1018 

\bibitem[{Anthony-Twarog {et~al.}(2013)}]{2013ApJ...767L..19A}Anthony-Twarog, B.~J., Deliyannis, C.~P., Rich, E., \& Twarog, B.~A. 2013, \apjl, 767, L19

\bibitem[Asplund et al.(2009)]{2009ARA&A..47..481A} Asplund, M., Grevesse, N., Sauval, A.~J., \& Scott, P.\ 2009, \araa, 47, 481 

\bibitem[{Basu {et~al.}(2011)}]{Basu:2011cc} Basu, S., Grundahl, F., Stello, D., et al.\ 2011, \apjl, 729, L10 

\bibitem[Bertelli et~al.(2008)]{Bertelli:2008ge}Bertelli, G., Girardi, L., Marigo, P., \& Nasi, E. 2008, \aap, 484, 815

\bibitem[Bertelli et~al.(2009)]{Bertelli:2009ic}Bertelli, G., Nasi, E., Girardi, L., \& Marigo, P. 2009, \aap, 508, 355

\bibitem[{Bessell {et~al.}(1998)}]{1998A&A...333..231B} Bessell, M.~S., Castelli, F., \& Plez, B. 1998, \aap,  333, 231

\bibitem[Boothroyd  \& Sackmann(1988)]{1988ApJ...328..641B} Boothroyd, A.~I., \& Sackmann, I.-J.\ 1988, \apj, 328, 641 

\bibitem[{Bragaglia {et~al.}(2001)}]{2001AJ....121..327B} Bragaglia, A., Carretta, E., Gratton, R.~G., et al.\ 2001, \aj, 121, 327 

\bibitem[Bragaglia \& Tosi(2006)]{2006AJ....131.1544B} Bragaglia, A., \& Tosi, M.\ 2006, \aj, 131, 1544 

\bibitem[{Carlberg {et~al.}(2012)}]{carlberg12} Carlberg, J.~K., Cunha, K., Smith, V.~V., \& Majewski, S.~R. 2012, \apj, 757, 109

\bibitem[Carlberg(2014)]{2014AJ....147..138C} Carlberg, J.~K.\ 2014, \aj, 147, 138 

\bibitem[Carr et al.(2000)]{2000ApJ...530..307C} Carr, J.~S., Sellgren, K., \& Balachandran, S.~C.\ 2000, \apj, 530, 307 


\bibitem[{Charbonnel \& Balachandran(2000)}]{Charbonnel:2000ud} Charbonnel, C. \& Balachandran, S.~C. 2000, \aap, 359,  563

\bibitem[Cunha et al.(2014)]{cunha14} Cunha, K., Smith, V.~V., Johnson, J.~A., et al.\ 2014, \apjl, in press

\bibitem[{Denissenkov(2012)}]{denissenkov12} Denissenkov, P.~A. 2012, \apjl, 753, L3

\bibitem[{Eggleton {et~al.}(2008)}]{eggleton08} Eggleton, P.~P., Dearborn, D. S.~P., \& Lattanzio, J.~C. 2008, \apj, 677, 581

\bibitem[Eisenstein et al.(2011)]{2011AJ....142...72E} Eisenstein, D.~J., Weinberg, D.~H., Agol, E., et al.\ 2011, \aj, 142, 72 

\bibitem[Gilroy(1989)]{gilroy89} Gilroy, K.~K.\ 1989, \apj, 347, 835 

\bibitem[{Gonz{\'a}lez~Hern{\'a}ndez \& Bonifacio(2009)}]{2009A&A...497..497G} Gonz{\'a}lez~Hern{\'a}ndez, J.~I. \& Bonifacio, P. 2009, \aap, 497, 497

\bibitem[Gunn et al.(2006)]{2006AJ....131.2332G} Gunn, J.~E., Siegmund,  W.~A., Mannery, E.~J., et al.\ 2006, \aj, 131, 2332 

\bibitem[Hill \& Pasquini(1999)]{1999A&A...348L..21H} Hill, V., \& Pasquini, L.\ 1999, \aap, 348, L21 

\bibitem[{Hole {et~al.}(2009)}]{TabethaHole:2009gi} Hole, K.~T., Geller, A.~M., Mathieu, R.~D., Platais, I., Meibom, S., \& Latham,   D.~W. 2009, \aj, 138, 159

\bibitem[Kjeldsen \& Bedding(1995)]{1995A&A...293...87K} Kjeldsen, H., \& Bedding, T.~R.\ 1995, \aap, 293, 87 

\bibitem[{Kumar {et~al.}(2011)}]{Kumar:2011jr} Kumar, Y.~B., Reddy, B.~E., \& Lambert, D.~L. 2011, \apj,  730, L12

\bibitem[Luck(1994)]{Luck:1994iu} Luck, R.~E.\ 1994, \apjs, 91, 309 

\bibitem[{Majewski(2012)}]{Majewski:2012wx} Majewski, S.~R. 2012, American Astronomical Society Meeting Abstracts {\#}219

\bibitem[{Marigo {et~al.}(2008)}]{Marigo:2008fy}Marigo, P., Girardi, L., Bressan, A., Groenewegen, M. A.~T., Silva, L., \&  Granato, G.~L. 2008, \aap, 482, 883

\bibitem[Meszaros et al.(2013)]{meszaros02} Meszaros, Sz., Holtzman, J., Garc{\'{\i}}a P{\'e}rez, A.~E. et al 2013, \aj, 146, 133

\bibitem[Mikolaitis et al.(2010)]{2010MNRAS.407.1866M} Mikolaitis, {\v S}., Tautvai{\v s}ien{\.e}, G., Gratton, R., Bragaglia, A., \& Carretta, E.\ 2010, \mnras, 407, 1866 

\bibitem[Mikolaitis et al.(2011a)]{2011MNRAS.413.2199M} Mikolaitis, {\v S}., Tautvai{\v s}ien{\.e}, G., Gratton, R., Bragaglia, A., \& Carretta, E.\ 2011a, \mnras, 413, 2199 

\bibitem[Mikolaitis et al.(2011b)]{2011MNRAS.416.1092M} Mikolaitis, {\v S}., Tautvai{\v s}ien{\.e}, G., Gratton, R., Bragaglia, A., \& Carretta, E.\ 2011b, \mnras, 416, 1092 

\bibitem[Mikolaitis et al.(2012)]{2012A&A...541A.137M} Mikolaitis, {\v S}., Tautvai{\v s}ien{\.e}, G., Gratton, R., Bragaglia, A., \& Carretta, E.\ 2012, \aap, 541, AA137 

\bibitem[Monaco et al.(2014)]{2014A&A...564L...6M} Monaco, L., Boffin, H.~M.~J., Bonifacio, P., et al.\ 2014, \aap, 564, LL6 

\bibitem[Nidever et al.(2015)]{2015arXiv150103742N} Nidever, D.~L., Holtzman, J.~A., Allende Prieto, C., et al.\ 2015, submitted to \aj, arXiv:1501.03742.


\bibitem[Paxton et al.(2013)]{2013ApJS..208....4P} Paxton, B., Cantiello, M., Arras, P., et al.\ 2013, \apjs, 208, 4 

\bibitem[Pinsonneault et al.(2014)]{pinsonneault14} Pinsonneault, M.~H., Elsworth, Y., Epstein, C., et al.\ 2014, \apjs, in~press

\bibitem[{Platais {et~al.}(2013)}]{Platais:2013im} Platais, I., Gosnell, N.~M., Meibom, S., Kozhurina-Platais, V., Bellini, A.,  Veillet, C., \& Burkhead, M.~S. 2013, \aj, 146, 43

\bibitem[Sanders(1972)]{sanders72} Sanders, W.~L.\ 1972, \aap, 19, 155 

\bibitem[Schlegel et al.(1998)]{1998ApJ...500..525S} Schlegel, D.~J., Finkbeiner, D.~P., \& Davis, M.\ 1998, \apj, 500, 525 

\bibitem[{Skrutskie {et~al.}(2006)}]{2006AJ....131.1163S} Skrutskie, M.~F., Cutri, R.~M., Stiening, R., et al.\ 2006, \aj, 131, 1163 

\bibitem[Smiljanic et al.(2009)]{Smiljanic:2009ik} Smiljanic, R., Gauderon, R., North, P., et al.\ 2009, \aap, 502, 267 

\bibitem[{Smith {et~al.}(2013)}]{2013ApJ...765...16S} Smith, V.~V., Cunha, K., Shetrone, M.~D., et al.\ 2013, \apj, 765, 16 

\bibitem[{Sneden(1973)}]{Sneden:1973el} Sneden, C. 1973, \apj, 184, 839

\bibitem[{Stello {et~al.}(2011)}]{Stello:2011hu} Stello, D., Meibom, S., Gilliland, R.~L., et al.\ 2011, \apj, 739, 13 

\bibitem[Tautvai{\v s}iene et~al.(2000)]{2000A&A...360..499T} Tautvai{\v s}iene, G., Edvardsson, B., Tuominen, I., \& Ilyin, I.\ 2000, \aap, 360, 499 

\bibitem[Tautvai{\v s}ien{\.e} et al.(2005)]{2005A&A...431..933T} Tautvai{\v s}ien{\.e}, G., Edvardsson, B., Puzeras, E., \& Ilyin, I.\ 2005, \aap, 431, 933 

\bibitem[{Wood {et~al.}(2014)Wood, Lawler, \& Shetrone}]{2014ApJ...787L..16W}Wood, M.~P., Lawler, J.~E., \& Shetrone, M.~D. 2014, \apjl, 787, L16

\bibitem[{Wright {et~al.}(2010)}]{2010AJ....140.1868W}Wright, E.~L., Eisenhardt, P.~R.~M., Mainzer, A.~K., et al.\ 2010, \aj, 140, 1868 

\bibitem[Zasowski et al.(2013)]{2013AJ....146...81Z} Zasowski, G., Johnson, J.~A., Frinchaboy, P.~M., et al.\ 2013, \aj, 146, 81 
%

\bibitem[Zucker \& Mazeh(1994)]{1994ApJ...420..806Z} Zucker, S., \& Mazeh, T.\ 1994, \apj, 420, 806 

\end{thebibliography}
\end{document}